# Synthesis and Microstructural Studies of Iron Based $LaO_{1-x}F_xFeAs$ Superconducting Materials

**Chandra Shekhar[1], Sonal Singh[1], P K Siwach[2], H K Singh[2] and O N Srivastava[1]**

[1]Department of Physics, Banaras Hindu University, Varanasi 221005, India

[2] National Physical Laboratory, Dr K S Krishnan Marg, New Delhi-10012, India

**Abstract**

We report on the synthesis and microstructural studies of fluorine based iron doped superconductors. We have successfully synthesized the fluorine based new superconducting material $LaO_{1-x}F_xFeAs$ by choosing comparatively lower temperature and longer synthesis duration. The superconducting transition temperature that we have achieved is 27.5 K which is observed at doping level of x=0.2. The structural microstructural characterizations have been done by employing XRD, SEM and TEM techniques. The SEM and TEM micrographs reveal the layered structure.

 Introduction

Since the discovery of superconductivity at 3.2K in iron-based compound [1], extensive efforts have been devoted to searching for new superconductors among this system. It has seen the height of day when a team led by Hosono at the Tokyo Institute of technology [Japan], replaced P atom by As atom together with substitution of oxygen with fluorine. The resultant compound $LaO_{1-x}F_xFeAs$ (x=0.11) shows the superconducting transition temperature ($T_C$) at 26 K [2]. Subsequently superconductivity at 25 K was also observed by partial substitution of La atom by Sr atom [3]. Shortly after this discovery, $T_C$ was surprisingly increased to more than 40 K when La in $LaO_{1-x}F_xFeAs$ was replaced by other rare earth elements such as Ce[4], Pr[5], Nd[6], Sm[7] and Gd [8].

The compound LaOFeAs is an equiatomic quaternary of ZrCuSiAs type tetragonal layered structure with lattice parameter *a*=4.035Å, *c*=8.739Å [9] and its structure belongs to the P4/nmm space group. The crystal is composed of a stack of alternating LaO and FeAs layers. The LaO layer is sandwiched between FeAs layers. It is thought that these two layers are, positively and negatively charged respectively, and that the La–O chemical bond in the LaO layer is ionic whereas the Fe–As has a predominantly covalent nature. Thus, the chemical formula may be expressed as $(La^{+3}O^{-2})^{+1}(FeAs)^{-1}$. The carriers has been increased by substitution of the $O^{-2}$ ion by $F^{-1}$ ion. The parent material LaOFeAs is non-superconducting but shows spin density wave instability in between 150−160K in both resistivity and d.c. magnetic susceptibility [2, 8]. The spin density wave instability relates some how to structural transition from tetragonal to monoclinic [10]. Doping the system with fluorine suppresses both the magnetic order and the structural distortion in favour of superconductivity. There have been only sparse studies of the microstructure of this new superconductor. Therefore, the emphasis in this communication is on exploration of microstructural characteristics of $LaO_{1-x}F_xFeAs$ superconductors

**Experimental Details**

In the present study the synthesis of F doped $LaO_{1-x}F_xFeAs$ (0≤ x ≤0.4) high temperature superconductor has been carried out by two step solid state reaction at ambient pressure. In the first step, for preparation of LaAs, $Fe_2As$ and FeAs, we mixed La (99.9% purity, 0.5−1 mm size, Leico), Fe (99.98% purity, 0.2−0.5 mm, Aldrich) and As (99.999% purity, Lump, Alfa-Aesar) in a ratio of 1:3:3 with the help of agate & pestle. The mixture powder was pelletized and then sealed in evacuated quarts tube in Ar atmosphere. The sealed silica tube was heated $900^0C$ for 12 hours. In the second step, the mixture of LaAs, $Fe_2As$ and FeAs was mixed with $La_2O_3$ (99.99% purity, 0.1−0.2 mm size, Aldrich), La and $LaF_3$ (99.9% purity, 0.1−0.2 mm size, Aldrich) in stoichiomentry ratio The final stoichiometry is $(1+x)La+(1-x)La_2O_3+xLaF_3+3FeAs$, x=0 for pure and x=0.05,0.1,0.2,0.3,0.4 for fluorine doped samples. After the final grinding, the powder was again pelletized at a pressure of 4tons/$inch^2$. The quartz tube was evacuated up to$10^{-5}$ torr and sealed. The sealed quartz tube was heated again at $1150^0C$ for 60 hours followed by furnace cooling to room temperature. We have chosen comparatively lower temperature and longer synthesis duration to avoid explosion. This is some what different than the standard synthesis temperature and duration so far adopted [7]. All the grindings have been carried out in a glove box containing $P_2O_5$, NaOH and under argon atmosphere. All the samples in the present investigation were subjected to gross structural characterization by X-ray diffraction (XRD, PANanalytical X'PRO, CuK$_\alpha$ radiation), electrical transport measurements by four probe technique (Keithley Resistivity-Hall setup), surface morphological characterization by scanning electron microscope (SEM, Philips XL-20), and the microstructural characterization by High Resolution Transmission electron microscope (TEM, FEI, Technai 20G$^2$).

**Results and Discussion**

Figure 1 show XRD patterns of LaOFeAs and LaO$_{0.8}$F$_{0.2}$FeAs samples. This XRD pattern reveals that the undoped and doped LaOFeAs samples are polycrystalline in nature. XRD analysis using a computerized program based on a least square fitting method gives lattice parameters *a*=4.039Å & *c* =8.742Å and *a*=4.030Å & *c* =8.716Å for pure LaOFeAs and doped (x=0.2) samples respectively. It is very close to the reported standard lattice parameter values [7, 9]. However these parameters are somewhat smaller than the reported standard values [7] The XRD patterns indicate that all samples have the standard of LaOFeAs structure with some minor impurity phases.

Figure 2 shows the resistance vs temperature behavior of pure and doped samples monitored by the standard four-probe method. The resistivity of LaOFeAs shows an anomaly at 155K, which is similar to that of other reports, where it has been shown to occur due to spin density wave instability [2, 6, 10]. For the sample LaO$_{0.8}$F$_{0.2}$FeAs, the T$_C$ is 27.5(±0.2) K. This value of T$_C$ is slightly higher in comparison to other report [2]. This may be explicable in terms of enhanced chemical pressure originated from shrinkage of lattice as brought out comparatively smaller lattice parameters of the phase synthesized in the present case.

The microstructural characterization of La$_{0.8}$F$_{0.2}$FeAs is shown by SEM micrographs (Figs. 3(a), 3(b) and (c)). One outstanding feature as depicted by the SEM picture is the presence of layered structure. Several layers forming a block can be easily seen in all the figures. These layered blocks are depicted by arrows in Figs. 3[a), 3(b) and (c)]. The layered blocks are shifted with respect to each other forming stair like structure. It appears that a group of [(La-O-F) (FeAs)] layers gets shifted with respect to the second such block. It suggests rather weak bonding between two [(La-O-F) (FeAs)] layers. These features are indicative of a layer growth mechanism for the formation of [(La-O-F) (FeAs)] phase.

Further details of microstructure are shown in the TEM micrographs as in Fig. 4(a). This confirms the layered nature of the [(LaO$_{0.8}$F$_{0.2}$) (FeAs)] phase. Some regions outlining the layered characteristics are depicted by arrows. Extensive explorations through Transmission Electron Microscopy revealed curious structural characteristics. This relates to the possible presence of supper lattice structure as manifested by the presence of extra spots (outlined by arrows) in the center of the main square grid of spots corresponding to the basic tetragonal phase. The indices the basic spot have been outlined. The presence of supper lattice spot suggests the existence of a tetragonal supper lattice with ½($\sqrt{2}$a) = a/$\sqrt{2}$ lattice parameter. Further studies of microstructural characteristics together with their possible correlation with superconducting behavior are being done and results will be forthcoming.

**Conclusions**

We have successfully synthesized the fluorine doped LaO$_{1-x}$F$_x$FeAs superconductors with transition temperature 27.5 K at doping level x=0.2. Microstructural characterization reveals that this new superconductors have layered structure with the existence of tetragonal superlattice structures.

**Acknowledgement**

The authors are grateful to Prof. A.R. Verma, Prof. C.N.R. Rao (FRS), Prof. D.P. Singh for their encouragement and Prof. H Kishan & Dr. V.P.S. Awana, NPL, New Delhi for fruitful discussion and suggestions. Financial supports from UGC and DST-UNANST are gratefully acknowledged.

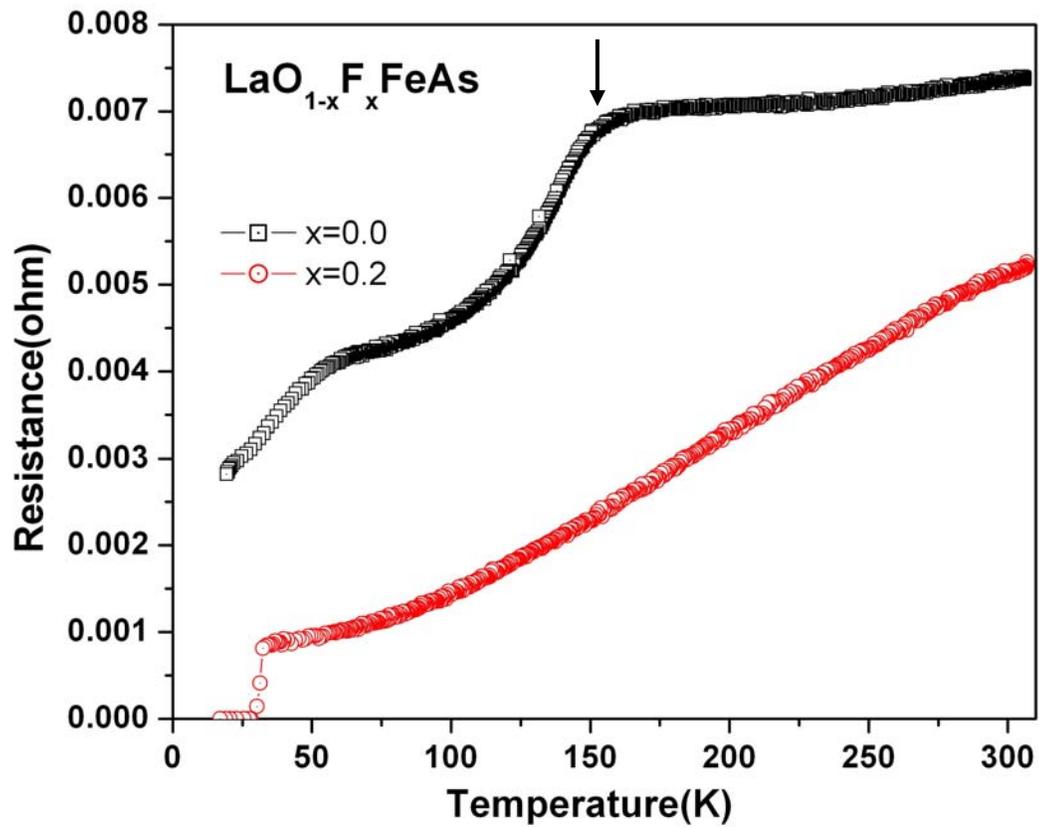

**Fig. 1** Resistance vs temperature behaviour of pure and fluorine doped LaOFeAs samples (lower curve). The superconducting transition temperature corresponds to 27.5K. The pristine sample exhibiting spin density wave anomaly marked by arrow but no superconducting transition is shown by upper curve

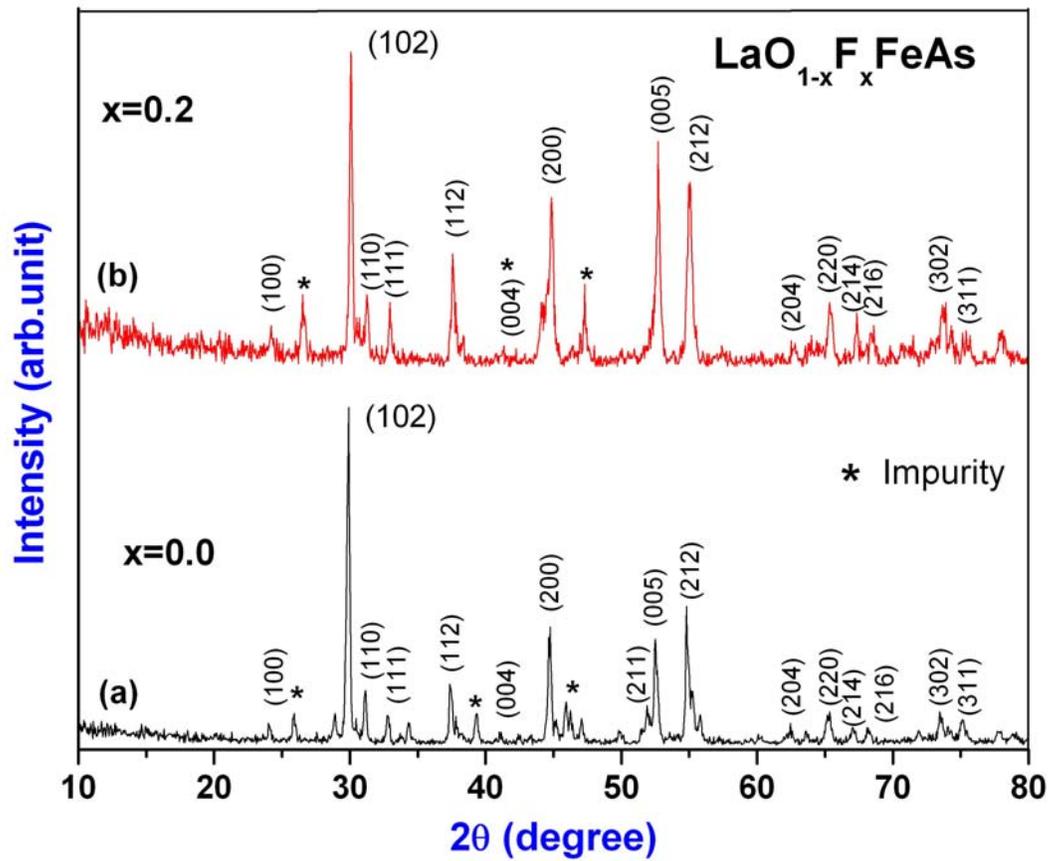

**Fig. 2** XRD pattern of pure and fluorine doped LaOFeAs samples

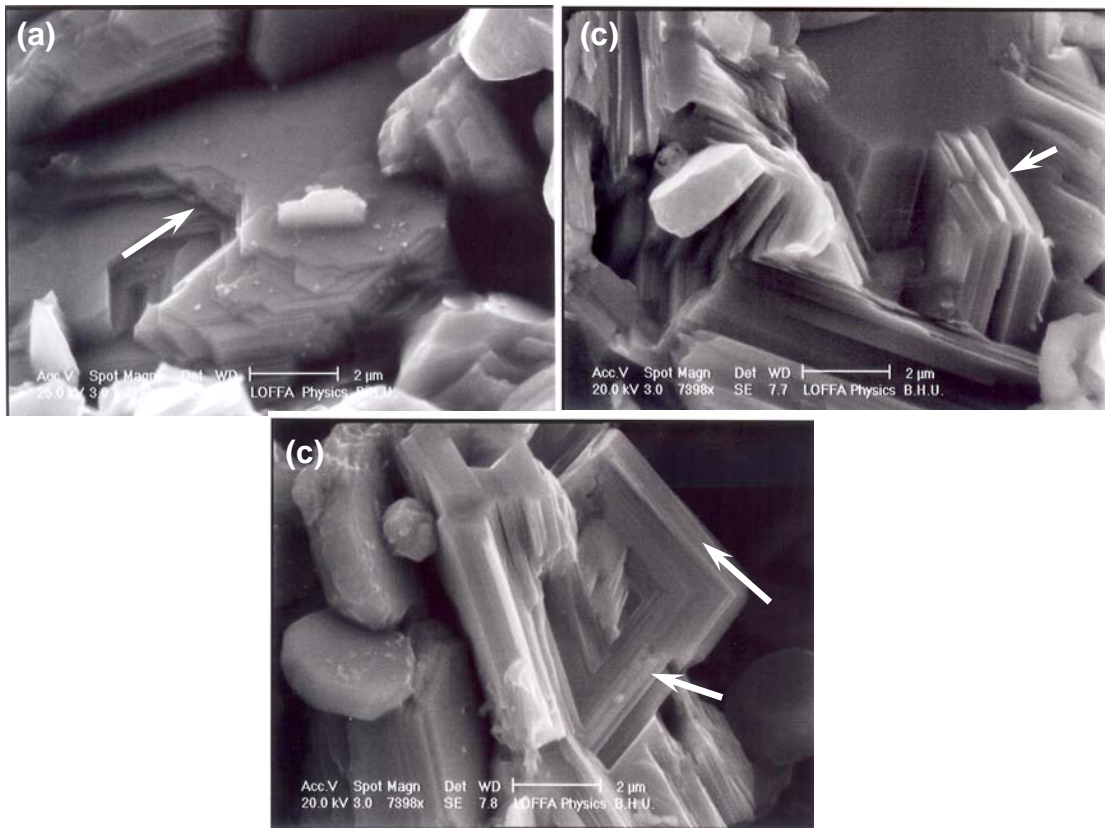

**Fig. 3** SEM micrographs of as synthesised $LaO_{1-x}F_xFeAs$ with x=0.2 fluorine doped sample. The presence of layers is indicated by arrows.

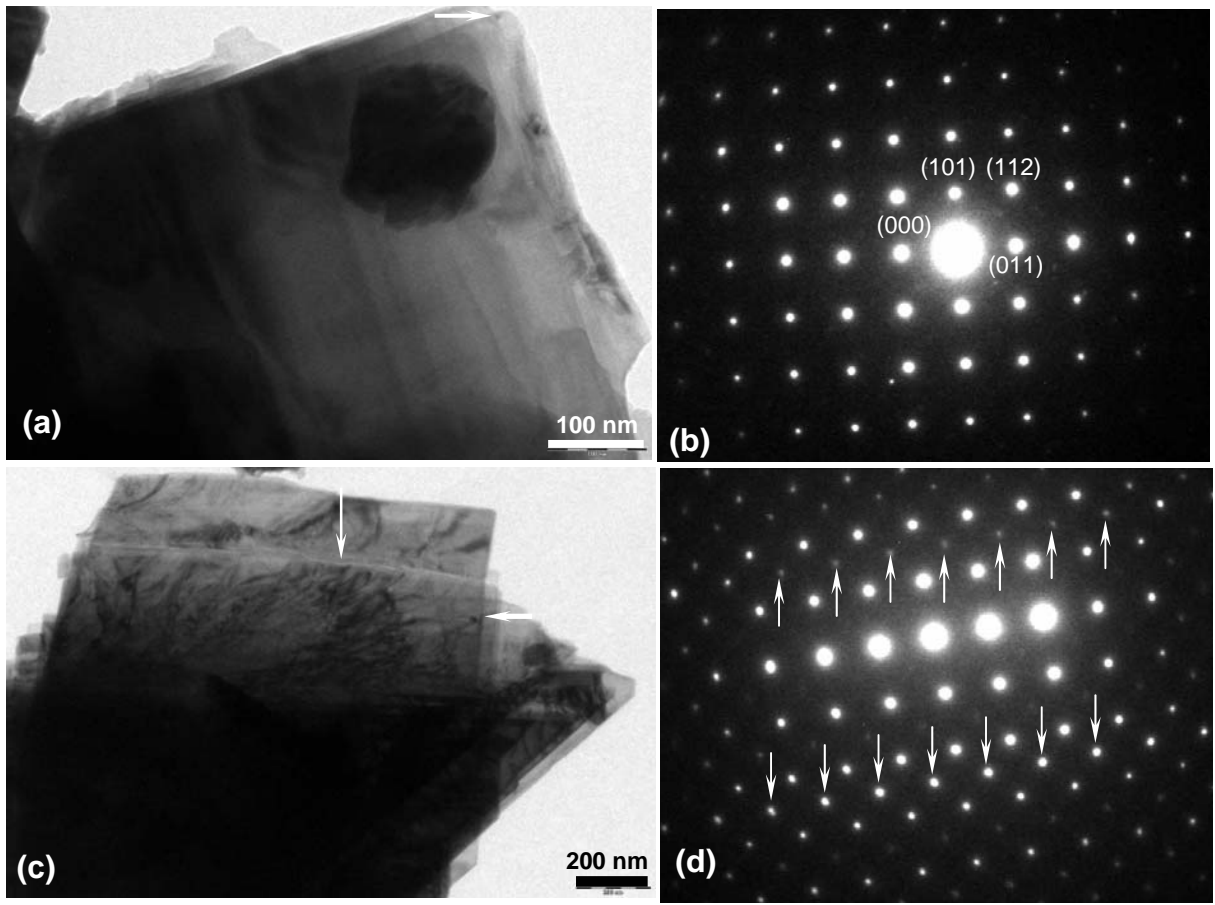

**Fig. 4** TEM micrographs of as synthesised LaO$_{1-x}$F$_x$FeAs with x=0.2 fluorine doped sample. The presence of layers in (a) & (b) is indicated by arrows. The presence of superlattice spots in (d) is indicated by arrows.